\documentclass{Interspeech2024}
\usepackage[dvipsnames]{xcolor}
\usepackage{multirow}
\usepackage{makecell}
\usepackage{pgfplots}
\usepackage{adjustbox}
\usepackage{pgfplotstable}
\usepackage{subcaption}
\usepackage{algorithm}
\usepackage{algpseudocode}
\usepackage{mathtools}
\usepackage{fixmath}
\usepackage{standalone}
\usepackage[multiple]{footmisc}
\usetikzlibrary{patterns.meta}

\graphicspath{{figures/}}

\DeclareMathOperator{\E}{\mathbb{E}}

\newcommand{\auxLoss}{Aux-Loss}
\newcommand{\simpleTopK}{\textit{Simple-Top-k}}
\newcommand{\iterativeZeroout}{\textit{Iterative-Zero-Out}}
\newcommand{\zeroNorm}{$L_0$-Norm}
\newcommand\blfootnote[1]{%
  \begingroup
  \renewcommand\thefootnote{}\footnote{#1}%
  \addtocounter{footnote}{-1}%
  \endgroup
}

\title{Dynamic Encoder Size Based on Data-Driven Layer-wise Pruning for Speech Recognition}

\name[affiliation={1,2,*}]{Jingjing}{Xu}
\name[affiliation={*, \dagger}]{Wei}{Zhou}
\name[affiliation={1,2}]{Zijian}{Yang}
\name[affiliation={2}]{Eugen}{Beck}
\name[affiliation={1,2}]{Ralf}{Schlüter}

\address{
  $^1$Machine Learning and Human Language Technology Group, Computer Science Dept., RWTH Aachen University, Germany\\
  $^2$AppTek GmbH, 52062 Aachen, Germany
  }
\email{\{jxu, zhou, zyang, schlueter\}@ml.rwth-aachen.de, ebeck@apptek.com}
\keywords{Speech recognition, Supernet training, Dynamic encoder, Pruning}

\renewrobustcmd{\bfseries}{\fontseries{b}\selectfont}
\renewrobustcmd{\boldmath}{}
\newrobustcmd{\B}{\bfseries}

\newsavebox\CBox

\makeatletter

\renewcommand{\section}{\@startsection
   {section}%
   {1}%
   {}%
   {-0.4\baselineskip}%
   {0.2\baselineskip}%
   {}}%

\renewcommand{\subsection}{\@startsection
  {subsection}%
  {2}%
  {}%
  {-0.1\baselineskip}%
  {0.1\baselineskip}%
  {}}%

\renewcommand{\subsubsection}{\@startsection
  {subsubsection}%
  {3}%
  {}%
  {-0.2\baselineskip}%
  {0.1\baselineskip}%
  {}}%

\begin{document}
\maketitle
\begin{abstract}
  Varying-size models are often required to deploy ASR systems under different hardware and/or application constraints such as memory and latency.
To avoid redundant training and optimization efforts for individual models of different sizes, we present the dynamic encoder size approach, which jointly trains multiple performant models within one supernet from scratch.
These subnets of various sizes are layer-wise pruned from the supernet, and thus, enjoy full parameter sharing.
By combining score-based pruning with supernet training, we propose two novel methods, {\simpleTopK} and {\iterativeZeroout}, to automatically select the best-performing subnets in a data-driven manner, avoiding resource-intensive search efforts.
Our experiments using CTC on both \textit{Librispeech} and \textit{TED-LIUM-v2} corpora show that our methods can achieve on-par performance as individually trained models of each size category.
Also, our approach consistently brings small performance improvements for the full-size supernet.
    \blfootnote{$^*$ denotes equal contribution}
    \blfootnote{$\dagger$ work done while at RWTH Aachen University, now at Meta}
\end{abstract}

\section{Introduction}

Automatic speech recognition (ASR) models run in different scenarios with different application needs and
computational budgets.
For some applications, inference speed is critical which often requires trading off
accuracy for model latency.
One of the simplest and most effective ways to reduce model latency is to reduce
model size.
For on-site ASR, edge devices have limited storage and memory budgets, thus
also imposing constraints on the model size.
Obtaining ASR models with different model sizes often requires optimizing training hyperparameters for each model individually.
However, repeated training results in high computational costs.
Therefore, arises the question of how to efficiently train models of
different sizes.

Pruning \cite{jiang2023microsoftpruning,yang2023l0norm,peng2023l0norm}, as a
model compression technique, is commonly used to obtain neural models of smaller sizes.
Pruning aims at removing unimportant weights from the network.
The lottery ticket hypothesis \cite{frankle2019lth,ding2022audiolottery}
discovers that there exists a sparse net in a full network that can achieve the
same performance.
However, pruning requires a converged base model and fine-tuning of each
small model separately, so the problem of repeated training remains unresolved.
The concept of supernet training is first proposed in
\cite{yu2019sandwichrule}.
The supernet and a fixed number of subnets fully share parameters and are
simultaneously trained.
After the joint training, all networks can achieve good convergence.
However, how to efficiently search for the subnets during training is still
challenging \cite{cai2020onceforall, chen2021autoformer}.

In this work, we combine the benefits of both ideas and demonstrate an efficient dynamic encoder training framework.
We leverage score-based layer-wise pruning to find the optimal layer combination
for the subnets, saving the computationally expensive search required by
the general supernet training methods \cite{yuan2023metaomni2, yang2022metaomni1}.
Furthermore, we design an efficient two-step training pipeline.
In Step 1, we propose two methods, {\simpleTopK} and {\iterativeZeroout}, to
effectively learn the associated layer importance scores in a data-driven way.
In step 2, we generate binary masks for all subnets and exploit the sandwich rule \cite{yu2019sandwichrule} for efficient joint training of the supernet and subnets.
Additionally, we explore different training techniques to mitigate the
mutual training inference and further boost the word error rate (WER).
We evaluate our approach by conducting experiments with the Conformer \cite{gulati2020conformer} connectionist temporal
classification (CTC) \cite{graves2006ctc} model on both \textit{Librispeech} and \textit{TED-LIUM-v2} datasets.
The results show that with our proposed framework, multiple models with the desired number of layers can be obtained in a single training job, each with competitive WER performance.
Even a slight WER improvement on the full-size model is obtained presumably due to the regularization effect of the co-trained subnets.
We also investigate the selected layers for the subnets and unexpectedly find
that the convolutional layers are selected the most.

\section{Related Work}
\subsection{Supernet/Subnet Joint Training}
\label{subsec:supernet_joint_train}
The RNN-T cascaded encoder architecture
\cite{narayanan2021cascaded,ding2022cascaded} utilizes the idea of auxiliary loss
\cite{tjandra2020auxloss,lee2021auxloss}, allowing direct connections between
intermediate encoder layers and decoders.
An advantage of this approach is that the training overhead is negligible
since no additional forward pass is required for the subnets.
Nevertheless, the low-level layers may not be the optimal choice for the
subnets.
With the same model size, there might be a better combination of layers  to
make up the subnets.
\cite{shi2021dynamictransducer} for example, comprises the subnets by choosing
every other or every third layer.
To avoid manual layer selection, our work utilizes a score-based pruning
method to achieve automatic layer selection during training.
\cite{rui2022lighthubert} randomly select one subnet from a total of 1000 subnets at each step of supernet training, which does not guarantee to find the optimal subnet under a specific size constraint.
\cite{yuan2023metaomni2, yang2022metaomni1} use evolutionary search to find the top-performing subnets under different
 size constraints from a predefined search space.
The WER and loss on the validation set are used as the ranking metric in
\cite{yuan2023metaomni2} and \cite{yang2022metaomni1}, respectively.
As a result, in one search procedure, all possible subnet candidates need to be
forwarded once with the whole validation data.
Although \cite{yuan2023metaomni2} leverages quantization to make the
inference more efficient,
repeating such a search procedure in each training step is computationally expensive.
\cite{fan2020structuredpruning,lee21pruning} use regularization tricks like stochastic depth \cite{huang2016stochasticdepth} and auxiliary loss \cite{tjandra2020auxloss,lee2021auxloss} to make model pruning aware and uses layer-wise pruning after training to search for subnets.
Compared to these methods, our work determines the optimal subnets based on
importance scores, thereby saving resource-intensive search efforts.
\cite{wu2021sparsity} employs unstructured gradient-based pruning criteria to
determine the subnets.
Unstructured pruning may result in the irregular sparse matrix which requires
reconstruction original dense shape in inference, hindering the acceleration
in practical use.
Thus, our work applies structured pruning to avoid irregular sparsity.

\subsection{Pruning}
Magnitude-based pruning \cite{song2015mp,cheng2021zeroout} preserves weights
with the highest absolute values.
However, the weight magnitude does not necessarily reflect the weight importance.
To address this issue, movement pruning \cite{victor2020movementpruning}
considers first-order information, which is how the weights change in training.
In this work, we follow their idea for our score-based pruning.
Furthermore, \cite{louizos2017l0nrom} introduces $L_0$ norm regularization on the
non-zero elements of weights, so that the models can be pruned to a
specified sparsity.
There has been growing interest in applying the $L_0$ norm for ASR tasks \cite{jiang2023microsoftpruning,yang2023l0norm,peng2023l0norm}.
We compare our approach with $L_0$ norm in Sec.
\ref{subsection:jointly_train_two_encoder}.

\section{Dynamic Encoder Size}
In this section, we present our approach to model encoders with dynamic size based on a supernet and M subnets that share parameters with the supernet.
Consider the learnable parameters $\theta = \{\theta_j\}_{j=1}^L$, where $L$ is the
total number of layers, $\theta_j$ denotes the parameters of layer $j$.
Let $\mathcal{C}=\{k_{1},...,k_m,...,k_M\}$ denotes a predefined set of expected number of layers for the M subnets $n_1, n_2,...,n_{M}$.
$\mathcal{C}$ is sorted in a decreasing order such that $k_1 = L$ and $k_M=k_{min}=\min_{\forall m\in M} k_m$.
For each subnet $n_m$, a binary pruning mask $\mathbold{z_m}\in\{0,1\}^L$ is learned such that $\sum_{j=1}^L z_m^j = k_m$, where $z_m^j$ denotes the mask for layer j.
$z_m^j=0$ indicates that layer $j$ is pruned for subnet $n_m$.
The layer importance scores are specified as $\mathbold{s}=\{s_l\}_{l=1}^L \in \mathbb{R}^L$.
The joint training optimization problem can be formulated as:
\scalebox{0.9}{\parbox{1.1\linewidth}{%
\begin{align*}
\min\limits_{\theta, \forall z_m} \E_{(x,y)\in \mathcal{D}} \left[ \mathcal{L}_{ASR}(x,y; \theta)\ + \sum_{m=1}^M \lambda_m \mathcal{L}_{ASR}(x,y; \theta \odot z_m)\right],
\end{align*}
}}
where $\mathcal{D}$ is the training data and $\lambda_m$ is the tunable loss scale, $\odot$ denotes the element-wise product.

We design an efficient 2-step training pipeline.
The main goal of Step 1 is to automatically learn the layer importance score.
In Step 1, the supernet and one subnet are jointly trained from
scratch.
The subnet is initialized with the full-size model and progressively layer-wise
pruned until its number of layers reaches $k_{min}$.
During this progressive pruning process, the layer scores
learned at each intermediate size category also reflecting the layer selection for the corresponding subnet.
We believe that with such a process, layer selection can be learned
in a smooth way.
Besides, the ASR loss of the full-sized supernet ensures that the performance of
the supernet is not compromised.
After the layer importance scores are learned, in Step 2, we generate binary
pruning masks for each subnet $n_1, n_2,...,n_{M}$ based on the importance score
 and then jointly train the supernet and all M subnets.


\subsection{Step 1 - Progressive Self-Pruning}
\label{sec:step1}
In Step 1, we progressively prune the subnet with a dynamically decreasing size, while the supernet and the subnet are jointly trained during the pruning process.
This step takes about 60\% of the total training time.
The subnet is initialized with 0 sparsity, which is defined as the percentage of pruned layers to the total number of layers.
We adopt an iterative training procedure to gradually increase the target sparsity of the subnet until it
reaches the desired maximum value $\frac{L-k_{min}}{L}$.
We denote I as the total number of training iterations, each with $\Delta T$ training steps.
In the $i$-th iteration, we set the number of layers of the subnet to
$k=L-\frac{(L-k_{min})\times i}{I}$.
In the following, we present two iterative self-pruning methods for the given k to learn the
associated layer importance.
\begin{figure}[t]
  \centering
  \includegraphics[scale=0.28]{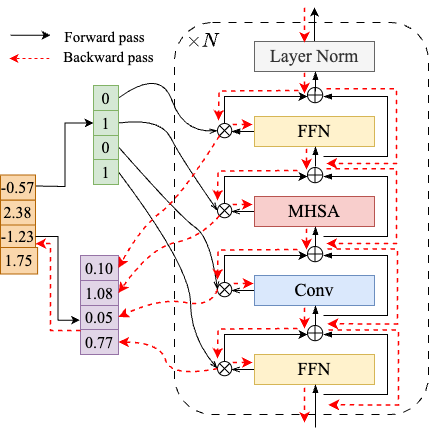}
  \caption{Illustration of {\simpleTopK}, STE uses a relaxed k-hot vector to estimate the gradients of the binary mask.}
  \label{fig:ste}
\end{figure}

\subsubsection{\simpleTopK}
{\simpleTopK} is a differentiable top-k operator inspired by \cite{victor2020movementpruning}.
For a given number of layers $k$, the pruning binary mask for the
subnet is $\mathbold{z}$ is $\{z^j | z^j = 1 \text{ if } s_j \text{ in } topk(\mathbold{s}, k) \text{ else } 0, j=1, 2, ..., $L$\}$.
Since such a binary mask $\mathbold{z}$ is not differentiable, {\simpleTopK} uses $\mathbold{z}$ in the forward pass to calculate the loss.
In the backward pass, as depicted in Figure \ref{fig:ste}, the straight through estimator (STE) \cite{yoshua2013ste} is
used to approximate the gradients for the step function.
We use a relaxed k-hot vector $\mathbold{\alpha}=[\alpha_1,...,\alpha_L]$ where $\sum_{j=1}^L \alpha_j = k, 0 \leq \alpha_j \leq 1$ to approximate the gradient of the binary mask $\mathbold{z}$.
The relaxed top-k algorithm is used to derive $\mathbold{\alpha}$ from $\mathbold{s}$, we refer the reader to \cite{sang2019subset} for more details
about the algorithm.

\subsubsection{\iterativeZeroout}
{\simpleTopK} uses STE, which leads to inconsistency in forward and backward passes.
Since the impact of the approximate gradient is unclear \cite{yin2019ste}, we design another method called {\iterativeZeroout}, to circumvent the usage of STE.
The pruning mask for the subnet here is defined as $\mathbold{z}=[f(s_1), f(s_2), ..., f(s_L)]$ where $f(.)$ is an activation function.
We use sigmoid for $f(.)$ in this work.
To constrain $\sum_{j=1}^L z^j = k$, we add an L1 norm term to the training
objective to ensure sparsity.
The total optimization loss can be formulated as \\
\scalebox{0.9}{\parbox{1.1\linewidth}{%
\begin{align*}
 \mathcal{L}_{ASR}(x,y; \theta)\ +  \lambda \mathcal{L}_{ASR}(x,y; \theta \odot \mathbold{z}) + \gamma \lvert\frac{\sum_{j=1}^{L}(z^j)-k}{L}\rvert,
\end{align*}
}}
where $\gamma$ is a tunable scale.
However, by simply adding a sparsity loss, we observe that $\mathbold{z}$ tends to
converge closer to a uniform distribution.
Therefore, we adopt the zero-out idea from \cite{cheng2021zeroout},
where each iteration adopts the following procedure:\\
1. Jointly train the supernet with parameters $\theta$ and the subnet with parameters
$\theta \odot \mathbold{z}$ for $\Delta T$ steps \\
2. Zero out the smallest $L-k$ elements in $\mathbold{z}$ but meanwhile keep them in the computation graph so that they can still get updated in further iterations and may have a chance to be revived.\\
In this way, the mask $\mathbold{z}$ will converge very close to a k-hot vector,
though not exactly.


\subsection{Step 2 - Supernet/Subnets Joint Training}
In Step 2, we set the binary mask $\mathbold{z_m}=\{z_m^j | z_m^j = 1 \text{ if } s_j \text{ in } topk(\mathbold{s}, k_m) \text{ else } 0\}$ for subnet $n_m$ with $k_m$ number of layers.
All binary masks are kept fixed in this step.
The sandwich rule, which is proposed in \cite{yu2019sandwichrule} and
successfully applied in \cite{yang2022metaomni1,yuan2023metaomni2}, is employed
in this step to improve efficiency.
More specifically, in each training step, we jointly
train the supernet, the smallest subnet, and one medium subnet randomly sampled from the remaining M-2 subnets.

Furthermore, we take advantage of layer dropout \cite{huang2016stochasticdepth}
to diminish the mutual interference between subnets and supernets during joint
training.
More precisely,  we apply layer dropout to those pruned layers that have indices
$\{j |  s_j \text{ not in } topk(\mathbold{s}, k_{min})\}$, similar to \cite{shi2021dynamictransducer}.
We also empirically observed that using around 40\% of the total training time is
already enough to reach good performance for all subnets.

\section{Experiments}
\subsection{Setup}
We conduct the experiments on the 960h \textit{Librispeech} corpus \cite{vassil2015lbs960} and
the \textit{TED-LIUM-v2} corpus \cite{rousseau2014tedlium}.
We use a phoneme-based CTC model as in \cite{zhou2023enhancing}.
We use a set of 79 end-of-word augmented phonemes \cite{wei2021phoneme}.
The acoustic model consists of a VGG front end and 12
Conformer \cite{gulati2020conformer} blocks.
In the Conformer block, we do not apply relative positional encoding.
Instead, we swap the order of the convolution module and the multi-head self-attention
module as in \cite{li2021swap} to speed up the training and inference.
The model size is set to 512 for \textit{Librispeech} and 384 for \textit{TED-LIUM-v2} corpus.
We use log Mel-filterbank features as input and specaug \cite{wei2020specaug} for data augmentation.
Similar to \cite{wei2022oclr}, one cycle learning rate scheduler is used for training
The learning rate (lr) is first linearly increased from $4\cdot10^{-6}$ to
$4\cdot10^{-4}$ for 45\% of
the training time, then linearly decreased from $4\cdot10^{-4}$ to $4\cdot10^{-6}$ for another 45\% of the
time.
For the rest 10\% of training time, the lr linearly decays to $10^{-7}$.
We train 50 epochs for \textit{TED-LIUM-v2} corpus and 30 epochs for \textit{Librispeech}.
The loss scale for each subnet in both training steps is set to 0.3.
In inference, we apply Viterbi decoding with a 4-gram word-level language model.
The config files and code to reproduce the results can be
found online\footnote{\scriptsize{\url{https://github.com/rwth-i6/returnn-experiments/tree/master/2024-dynamic-encoder-size}}}.

\subsection{1 Supernet + 1 Subnet}
\label{subsection:jointly_train_two_encoder}
In Table \ref{tab:jointly_train_two_model}, we compare ASR results of encoders
 with 24 and 48 layers trained using different methods.
For the \textit{\auxLoss} method, we add an auxiliary CTC loss with a loss scale of 0.3 to
the output of the 6-th Conformer block (corresponds to 24 layers).
We can see that for the 48 layers, both {\simpleTopK} and
\textit{\iterativeZeroout} achieve the same or slightly better WER.
Compared to the individually trained model, {\simpleTopK} and
\textit{\iterativeZeroout} achieve on-par WER for the 48-layer model and better WER
for the 24-layer model.
Compared to \textit{\auxLoss}, both {\simpleTopK} and
{\iterativeZeroout} gain slight WER improvement on the 48-layer model and
substantially outperform the 24-layer model by a relative $\sim$8\% with even fewer
parameters.
It confirms our hypothesis in Sec. \ref{subsec:supernet_joint_train}
that low-level layers are not the optimal choice
for the subnet.

We also compare our proposed method with the widely-used pruning method \textit{\zeroNorm} \cite{louizos2017l0nrom}.
We use the implementation from \cite{xia2022l0norm}.
We adapt \textit{\zeroNorm} to our joint training scenario and apply it to
the layer-wise self-pruning in Step 1.
The results in Table \ref{tab:jointly_train_two_model} show that
there is a considerable WER degradation for the 24-layer model trained from
\textit{\zeroNorm} compared to other methods.
The reason could be that the mask $\mathbold{z}$ is generated by
adding a random variable u $\sim U(0,1)$ distribution.
At each training step, the layer outputs of the subnet are
scaled by a fluctuating variable, which may cause disturbance to training.

\begin{table}[t]
  \centering
  \caption{\textit{ASR results comparison between different approaches for training two encoders with 48 and 24 layers on \textit{TED-LIUM-v2} dev set. The 48-layer model has a total of 41.7 M parameters.}}
  \label{tab:jointly_train_two_model}
  \setlength{\tabcolsep}{0.3em}
  \scalebox{0.95}{\begin{tabular}{|l|c||c|c|}
  \hline
  \multirow{2}{*}{Training} & Large & \multicolumn{2}{c|}{Small} \\
  & WER[\%] & \shortstack{Params. $[\text{M}]$}  & \shortstack{WER[\%]} \\ \hline

  separately & 7.5 & \multirow{2}{*}{20.9} & 8.4 \\ \cline{1-2} \cline{4-4}
  \auxLoss & 7.6 & & 8.8 \\ \hline
  \zeroNorm & 7.7 & 18.7 & 9.5\\ \hline
  \simpleTopK & 7.5 & 19.7 & \textbf{8.1} \\ \hline
  \iterativeZeroout & \textbf{7.4} & 18.8 & 8.2\\ \hline
\end{tabular}}
\end{table}

\subsection{1 Supernet + 2 Subnets}
Table \ref{tab:tedlium_three_model} and Table \ref{tab:lbs_three_model} report the
WER of models with 16, 32, and 48 layers on \textit{TED-LIUM-v2} and
\textit{Librispeech} test set.
On \textit{TED-LIUM-v2} test set, {\simpleTopK} performs best across
all three model sizes.
For the \textit{Librispeech} test set, {\simpleTopK} performs best
only on large and medium-sized models.
Both the proposed methods outperform the separately trained baselines for large and medium models, while performance degradation is observed for the small one.
A likely reason is that the loss scale for supernet is 1 and for all subnets is 0.3, thus placing more emphasis on the supernet during training.
Furthermore, we observe that the joint training can improve the WER of the supernet.
One possible explanation is that the layer masking on the shared
parameters introduces some regularization effect, similar to LayerDrop \cite{fan2020structuredpruning}.
In addition, the models trained from {\iterativeZeroout} perform slightly
worse than {\simpleTopK}.


\begin{table}[ht]
  \centering
  \caption{\textit{ASR results of three encoders with 48, 32, and 16 layers on
  \textit{TED-LIUM-v2} test set. The 48-layer model has a total of 41.7 M parameters.}}
  \label{tab:tedlium_three_model}
  \setlength{\tabcolsep}{0.3em}
  \scalebox{0.78}{\begin{tabular}{|l|c||c|c||c|c|}
  \hline
  \multirow{2}{*}{Training} & Large & \multicolumn{2}{c||}{Medium} &  \multicolumn{2}{c|}{Small} \\  \cline{2-6}
  & WER[\%] & Params.$[\text{M}]$ & WER[\%] & Params.$[\text{M}]$ & WER[\%] \\ \hline

  separately & 8.1 & \multirow{2}{*}{28.1} & 8.4 & \multirow{2}{*}{14.4} & 9.3\\ \cline{1-2}\cline{4-4}\cline{6-6}
  \auxLoss & 7.9 & & 8.6 & & 10.9\\ \hline
  \simpleTopK & \textbf{7.8} & 27.5 & \textbf{8.0} & 14.1 & \textbf{9.1}\\ \hline
  \iterativeZeroout & 8.1 & 25.8 & 8.2 & 12.9 & 9.6\\ \hline
\end{tabular}}
\end{table}

\begin{table}[ht]
  \centering
  \caption{\textit{ASR results of three encoders with 48, 32 and 16 layers on
  \textit{Librispeech} test set. The 48-layer model has a total of 74.1 M parameters.}}
  \label{tab:lbs_three_model}
  \setlength{\tabcolsep}{0.3em}
  \scalebox{0.78}{\begin{tabular}{|l|c|c||c|c|c||c|c|c|}
  \hline
  \multirow{3}{*}{Training} & \multicolumn{2}{c||}{Large} & \multicolumn{3}{c||}{Medium}
  & \multicolumn{3}{c|}{Small} \\ \cline{2-9}
  & \multicolumn{2}{c||}{WER[\%]} & Params. & \multicolumn{2}{c||}{WER[\%]}
  & Params. & \multicolumn{2}{c|}{WER[\%]}
  \\ \cline{2-3} \cline{5-6} \cline{8-9}
  & clean & other & $[\text{M}]$ & clean & other & $[\text{M}]$ & clean & other
  \\ \hline
  separately & 3.3 & 7.1 & \multirow{2}{*}{49.9} & 3.5 & 7.7 &
  \multirow{2}{*}{25.6} & \textbf{3.6} & \textbf{8.4} \\ \cline{1-3} \cline{5-6} \cline{8-9}
  \auxLoss & 3.2 & 6.9 & & 3.6 & 7.9 & & 4.5 & 9.7\\ \hline
  \simpleTopK & \textbf{3.1} & \textbf{6.8} & 47.1 & \textbf{3.2} & \textbf{7.0} & 23.8 & 3.9 & 9.1 \\ \hline
  \iterativeZeroout & 3.2 & 7.1 & 47.0 & \textbf{3.2} & 7.2 & 26.7 & 4.1 & 9.6 \\ \hline
\end{tabular}}
\end{table}

\subsection{Ablation Study}
\subsubsection{Layer Dropout}
The layer dropout in Step 2 can alleviate the mutual interference between the supernet
and subnets, thus playing an important role in joint training.
We explicitly study the impact of the layer dropout and present the result in
Table \ref{tab:layer_dropout_study}.
If layer dropout is applied in Step 1, we only apply it on the unselected layers, i.e., $z^j \neq 1$.
Table \ref{tab:layer_dropout_study} demonstrates that it is not necessary to
apply layer dropout in Step 1.
The best result is achieved by only applying layer dropout with a value of 0.3
in Step 2.
Additionally, we have also tried to apply dropout on the entire layer group as in \cite{fan2020structuredpruning}.
However, we empirically find out that applying dropout to each layer individually
performs better.
This may make the training more robust since in each training step, different
combinations of layers can be dropped out.

\begin{table}[t]
  \centering
  \caption{\textit{ASR results of applying different values of layer dropout
  in Step 1 and Step 2 on \textit{Librispeech} test set. Method {\simpleTopK}
  is used to train three models with 16,32,48 layers respectively.}}
  \label{tab:layer_dropout_study}
  \setlength{\tabcolsep}{0.3em}
  \scalebox{0.9}{\begin{tabular}{|c|c||c|c||c|c||c|c|}
  \hline
  \multicolumn{2}{|c|}{\multirow{2}{*}{Layer dropout}} & \multicolumn{6}{c|}{WER [\%]}\\ \cline{3-8}
  \multicolumn{2}{|c|}{} & \multicolumn{2}{c||}{Large} & \multicolumn{2}{c||}{Medium} &  \multicolumn{2}{c|}{Small} \\ \hline
  Step 1 & Step 2 & clean & other  & clean & other  & clean & other \\ \hline
  \multirow{4}{*}{n/a} & 0 & \textbf{3.1} & 6.9 & \textbf{3.2} & 7.1 & 4.1 & 9.7  \\
  \cline{2-8}
  & 0.1 & 3.2 & \textbf{6.8} & \textbf{3.2} & \textbf{7.0} & 4.2 & 9.8 \\ \cline{2-8}
  & 0.3 & \textbf{3.1} & \textbf{6.8} & \textbf{3.2} & \textbf{7.0} & \textbf{3.9} & \textbf{9.1}
  \\ \cline{2-8}
  & 0.5 & 3.3 & 7.2 & 3.3 & 7.3 & 4.1 & 9.5 \\ \hline
  0.1 & \multirow{2}{*}{0.3} & 3.3 & 7.0 & \textbf{3.2} & 7.2 & 4.2 & 9.6 \\
  \cline{1-1} \cline{3-8}
  0.3 & & 3.2 & 7.0 & \textbf{3.2} & \textbf{7.0} & 4.0 & 9.2\\ \hline
\end{tabular}}
\end{table}

\subsubsection{Number of Self-Pruning Iterations}
Table \ref{tab:num_prun_iters} compares the WERs of employing a different number of pruning iterations in Step 1 in Sec. \ref{sec:step1}.
The WERs of all three size models tend to decrease when more
pruning iterations are used.
As the models are trained from scratch in Step 1, the layer importance may
change significantly during training.
Presumably, using more iterations avoids making suboptimal decisions that select suboptimal layers in the early stages.

\begin{table}[ht]
  \centering
  \caption{\textit{ASR results of employing different number of self-pruning
  iterations in Step 1 on \textit{Librispeech} test set. {\simpleTopK}
  is used to train three models with 16,32,48 layers respectively.}}
  \label{tab:num_prun_iters}
  \setlength{\tabcolsep}{0.3em}
  \scalebox{0.95}{\begin{tabular}{|c||c|c||c|c||c|c|}
  \hline
  \multirow{3}{*}{\# iterations} & \multicolumn{6}{c|}{WER [\%]} \\ \cline{2-7}
   & \multicolumn{2}{c||}{Large} & \multicolumn{2}{c||}{Medium} &  \multicolumn{2}{c|}{Small} \\ \cline{2-7}
  & clean & other & clean & other & clean & other \\ \hline
  1 & 3.3 & 7.1 & 3.4 & 7.3 & \textbf{3.8} & 9.3 \\ \hline
  2 & 3.2 & 7.1 & \textbf{3.2} & 7.2 & 3.9 & 9.5 \\ \hline
  4 & 3.3 & 7.0 & 3.3 & 7.2 & 4.1 & 9.3   \\ \hline
  8 & \textbf{3.1} & 6.9 & \textbf{3.2} & 7.2 & 3.9 & 9.2\\ \hline
  32 & \textbf{3.1} & \textbf{6.8} & \textbf{3.2} & \textbf{7.0} & 3.9
   & \textbf{9.1} \\ \hline
\end{tabular}}
\end{table}

\subsubsection{Training Time Distribution}
Table \ref{tab:time_distribution} shows the WER results of using different training time distributions in Step 1 and Step 2.
We observe that using 60\% of training time in Step 1 and 40\% in Step 2 achieves the best performance.
If less time is distributed to Step 1, the layer importance scores may not be learned well, leading to premature decisions.
On the contrary, allocating more time in Step 1 will lead to insufficient joint training time under a fixed training budget and also lead to performance degradation.

\begin{table}[ht]
  \centering
  \caption{\textit{ASR results of using different training
  time distribution in Step 1 and Step 2 on \textit{Librispeech} test set. Method {\simpleTopK}
  is used to train three models with 16,32,48 layers respectively.}}
  \label{tab:time_distribution}
  \setlength{\tabcolsep}{0.3em}
  \scalebox{1}{\begin{tabular}{|c|c||c|c||c|c||c|c|}
  \hline
  \multirow{3}{*}{Step 1} & \multirow{3}{*}{Step 2} & \multicolumn{6}{c|}{WER[\%]} \\ \cline{3-8}
  & & \multicolumn{2}{c||}{Large} & \multicolumn{2}{c||}{Medium} &  \multicolumn{2}{c|}{Small} \\ \cline{3-8}
  & & clean & other  & clean & other  & clean & other \\ \hline
  50\% & 50\% & 3.3 & 7.1 & 3.3 & 7.2 & 4.0 & 9.3 \\ \hline
  60\% & 40\% & \textbf{3.1} & \textbf{6.8} & \textbf{3.2} & \textbf{7.0} & \textbf{3.9} & \textbf{9.1} \\ \hline
  70\% & 30\% & \textbf{3.1} & 7.1 & \textbf{3.2} & 7.2 & 4.4 & 10.9 \\ \hline
\end{tabular}}
\end{table}

\subsection{Sandwich Rule: 1 Supernet and M$>$2 Subnets}
Table \ref{tab:tedlium_four_model} shows the effectiveness of applying the sandwich
rule.
Compared to jointly training three models, there is no increase in
the training computation.
In Step 2, we only need to update the 48-layer
model, 12-layer model, and one randomly selected model with 24 or 36 layers.
Yet we can obtain one more subnet and the WERs of all four models
are still competitive.

\begin{figure}[t!]
    \centering
    \subfloat[\simpleTopK-36]{
    \pgfplotstableread{
Label FFN1 MHSA Conv FFN2
1-3 2 3 3 2
4-6 1 2 3 3
7-9 1 1 3 3
10-12 3 0 3 3
}\simpleTopK

\begin{tikzpicture}[scale=0.33]
  \tikzstyle{every node}=[font=\large]
    \begin{axis}[
        ybar stacked,
        ymin=0,
        ymax=12,
        xtick=data,
        legend columns=-1,
        legend style={at={(0,1.1)},nodes={scale=1.23, transform shape}, anchor=west},
        reverse legend=true,
        xticklabels from table={\simpleTopK}{Label},
        xticklabel style={text width=1cm,align=center},
    ]
        \addplot [fill=YellowGreen!30]
            table [y=FFN1, meta=Label, x expr=\coordindex]
                {\simpleTopK};
                    \addlegendentry{FFN1}
        \addplot [fill=Tan!30!, postaction={pattern={Lines[angle=0, distance=1.5mm, line width=0.03mm]}}]
            table [y=MHSA, meta=Label, x expr=\coordindex]
                {\simpleTopK};
                    \addlegendentry{MHSA}
        \addplot [fill=NavyBlue!30]
            table [y=Conv, meta=Label, x expr=\coordindex]
                {\simpleTopK};
                    \addlegendentry{Conv}
        \addplot [fill=Dandelion!30,postaction={pattern={Lines[angle=45, distance=1mm, line width=0.03mm]}},nodes near coords,point meta=y]
            table [y=FFN2, meta=Label, x expr=\coordindex]
                {\simpleTopK};
                \addlegendentry{FFN2}

    \end{axis}
\end{tikzpicture}
    }
    \subfloat[\simpleTopK-24]{
    \pgfplotstableread{
Label FFN1 MHSA Conv FFN2
1-3 1 1 2 2
4-6 1 0 3 1
7-9 0 0 3 1
10-12 3 0 3 3
}\data

\begin{tikzpicture}[scale=0.33]
  \tikzstyle{every node}=[font=\large]
    \begin{axis}[
        ybar stacked,
        ymin=0,
        ymax=12,
        xtick=data,
        legend columns=-1,
        legend style={at={(0.5,1.2)},anchor=west},
        reverse legend=true,
        xticklabels from table={\data}{Label},
        xticklabel style={text width=1cm,align=center},
    ]
        \addplot [fill=YellowGreen!30]
            table [y=FFN1, meta=Label, x expr=\coordindex]
                {\data};
        \addplot [fill=Tan!30!, postaction={pattern={Lines[angle=0, distance=1.5mm, line width=0.03mm]}}]
            table [y=MHSA, meta=Label, x expr=\coordindex]
                {\data};
        \addplot [fill=NavyBlue!30]
            table [y=Conv, meta=Label, x expr=\coordindex]
                {\data};
        \addplot [fill=Dandelion!30,postaction={pattern={Lines[angle=45, distance=1mm, line width=0.03mm]}},nodes near coords,point meta=y]
            table [y=FFN2, meta=Label, x expr=\coordindex]
                {\data};

    \end{axis}
\end{tikzpicture}
    }      
    \subfloat[\simpleTopK-12]{
      \pgfplotstableread{
Label FFN1 MHSA Conv FFN2
1-3 1 1 2 1
4-6 0 0 1 0
7-9 0 0 2 1
10-12 2 0 1 0
}\simpleTopK

\begin{tikzpicture}[scale=0.33]
  \tikzstyle{every node}=[font=\large]
    \begin{axis}[
        ybar stacked,
        ymin=0,
        ymax=12,
        xtick=data,
        legend style={at={(1.05,0.5)},anchor=west},
        reverse legend=true,
        xticklabels from table={\simpleTopK}{Label},
        xticklabel style={text width=1cm,align=center},
    ]
        \addplot [fill=YellowGreen!30]
            table [y=FFN1, meta=Label, x expr=\coordindex]
                {\simpleTopK};
        \addplot [fill=Tan!30!, postaction={pattern={Lines[angle=0, distance=1.5mm, line width=0.03mm]}}]
            table [y=MHSA, meta=Label, x expr=\coordindex]
                {\simpleTopK};
        \addplot [fill=NavyBlue!30]
            table [y=Conv, meta=Label, x expr=\coordindex]
                {\simpleTopK};
        \addplot [fill=Dandelion!30,postaction={pattern={Lines[angle=45, distance=1mm, line width=0.03mm]}},nodes near coords,point meta=y]
            table [y=FFN2, meta=Label, x expr=\coordindex]
                {\simpleTopK};

    \end{axis}
\end{tikzpicture}
    }
    \\
    \subfloat[Iter-Zero-Out-36]{
      \pgfplotstableread{
Label FFN1 MHSA Conv FFN2
1-3 1 3 3 2
4-6 1 3 3 2
7-9 0 1 3 3
10-12 3 2 3 3
}\zerooutMedium

\begin{tikzpicture}[scale=0.33]
  \tikzstyle{every node}=[font=\large]
    \begin{axis}[
        ybar stacked,
        ymin=0,
        ymax=12,
        xtick=data,
        legend columns=-1,
        legend style={at={(0.5,1.2)},anchor=west},
        reverse legend=true,
        xticklabels from table={\zerooutMedium}{Label},
        xticklabel style={text width=1cm,align=center},
    ]
        \addplot [fill=YellowGreen!30]
            table [y=FFN1, meta=Label, x expr=\coordindex]
                {\zerooutMedium};
        \addplot [fill=Tan!30!, postaction={pattern={Lines[angle=0, distance=1.5mm, line width=0.03mm]}}]
            table [y=MHSA, meta=Label, x expr=\coordindex]
                {\zerooutMedium};
        \addplot [fill=NavyBlue!30]
            table [y=Conv, meta=Label, x expr=\coordindex]
                {\zerooutMedium};
        \addplot [fill=Dandelion!30,postaction={pattern={Lines[angle=45, distance=1mm, line width=0.03mm]}},nodes near coords,point meta=y]
            table [y=FFN2, meta=Label, x expr=\coordindex]
                {\zerooutMedium};

    \end{axis}
\end{tikzpicture}
    }
    \subfloat[Iter-Zero-Out-24]{
      \pgfplotstableread{
Label FFN1 MHSA Conv FFN2
1-3 1 1 2 0
4-6 0 1 3 2
7-9 0 0 3 3
10-12 1 1 3 3
}\zerooutMedium

\begin{tikzpicture}[scale=0.33]
  \tikzstyle{every node}=[font=\large]
    \begin{axis}[
        ybar stacked,
        ymin=0,
        ymax=12,
        xtick=data,
        legend columns=-1,
        legend style={at={(0.5,1.2)},anchor=west},
        reverse legend=true,
        xticklabels from table={\zerooutMedium}{Label},
        xticklabel style={text width=1cm,align=center},
    ]
        \addplot [fill=YellowGreen!30]
            table [y=FFN1, meta=Label, x expr=\coordindex]
                {\zerooutMedium};
        \addplot [fill=Tan!30!, postaction={pattern={Lines[angle=0, distance=1.5mm, line width=0.03mm]}}]
            table [y=MHSA, meta=Label, x expr=\coordindex]
                {\zerooutMedium};
        \addplot [fill=NavyBlue!30]
            table [y=Conv, meta=Label, x expr=\coordindex]
                {\zerooutMedium};
        \addplot [fill=Dandelion!30,postaction={pattern={Lines[angle=45, distance=1mm, line width=0.03mm]}},nodes near coords,point meta=y]
            table [y=FFN2, meta=Label, x expr=\coordindex]
                {\zerooutMedium};

    \end{axis}
\end{tikzpicture}
    }      
    \subfloat[Iter-Zero-Out-12]{
      \pgfplotstableread{
Label FFN1 MHSA Conv FFN2
1-3 1 1 2 0
4-6 0 0 0 1
7-9 0 0 2 3
10-12 1 0 1 0
}\zerooutMedium

\begin{tikzpicture}[scale=0.33]
  \tikzstyle{every node}=[font=\large]
    \begin{axis}[
        ybar stacked,
        ymin=0,
        ymax=12,
        xtick=data,
        legend columns=-1,
        legend style={at={(0.5,1.2)},anchor=west},
        reverse legend=true,
        xticklabels from table={\zerooutMedium}{Label},
        xticklabel style={text width=1cm,align=center},
    ]
        \addplot [fill=YellowGreen!30]
            table [y=FFN1, meta=Label, x expr=\coordindex]
                {\zerooutMedium};
        \addplot [fill=Tan!30!, postaction={pattern={Lines[angle=0, distance=1.5mm, line width=0.03mm]}}]
            table [y=MHSA, meta=Label, x expr=\coordindex]
                {\zerooutMedium};
        \addplot [fill=NavyBlue!30]
            table [y=Conv, meta=Label, x expr=\coordindex]
                {\zerooutMedium};
        \addplot [fill=Dandelion!30,postaction={pattern={Lines[angle=45, distance=1mm, line width=0.03mm]}},nodes near coords,point meta=y]
            table [y=FFN2, meta=Label, x expr=\coordindex]
                {\zerooutMedium};

    \end{axis}
\end{tikzpicture}
    }
    \caption{The distribution of selected layers for the models with 12, 24 and
    36 layers shown in Table \ref{tab:tedlium_four_model}.}
    \label{Fig:layers_analysis}
\end{figure}
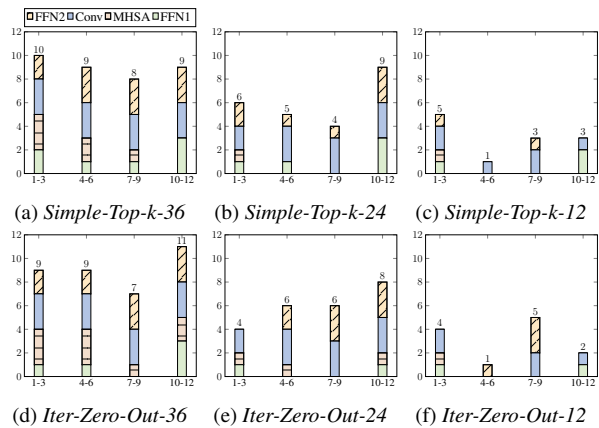

We further analyse the selected layers for the models with 12, 24, and 36 layers in
Figure \ref{Fig:layers_analysis}.
We note that the feed-forward layers tend to be pruned the most, followed by
multi head self-attention (MHSA) layers.
Also, the remaining MHSA layers tend to be distributed at
the bottom layer.
To our surprise, the convolutional layers are the most
selected layers, indicating that it is generally more important for
our phoneme CTC model.

\begin{table}[t]
  \centering
  \caption{\textit{ASR results of four encoders with 12, 24, 36 and 48 layers trained using the sandwich rule on
  \textit{TED-LIUM-v2} test set. The 48-layer model has totally 41.7 M parameters.}}
  \label{tab:tedlium_four_model}
  \setlength{\tabcolsep}{0.3em}
  \scalebox{0.75}{\begin{tabular}{|l|c||c|c||c|c||c|c|c|}
  \hline
  \multirow{2}{*}{Training} & Large & \multicolumn{2}{c||}{Medium 1} & \multicolumn{2}{c||}{Medium 2} &  \multicolumn{2}{c|}{Small} \\  \cline{2-8}
  & \shortstack{WER\\$[\%]$}
  & \shortstack{Params. \\ $[\text{M}]$}  & \shortstack{WER\\$[\%]$}
  & \shortstack{Params. \\ $[\text{M}]$}  & \shortstack{WER\\$[\%]$}
  & \shortstack{Params. \\ $[\text{M}]$}  & \shortstack{WER\\$[\%]$}\\ \hline
  separately & 8.1 & 31.4 & 8.3 & 21.2 & \textbf{8.6} & 10.9 & \textbf{9.8}\\ \hline
  \auxLoss & 8.1 & 31.4 & 8.3 & 21.2 & 9.1 & 10.9 & 12.2 \\ \hline
  \simpleTopK & 8.0 & 31.1 & \textbf{8.1} & 20.6 & \textbf{8.6} & 10.0 & 10.2 \\ \hline
  \iterativeZeroout & \textbf{7.9} & 29.3 & \textbf{8.1} & 19.3 & 8.8 & 10.7 & 10.3 \\ \hline
\end{tabular}}
\end{table}

\section{Conclusion}
In this work, we present an efficient training scheme to obtain models of various sizes by combining score-based
model pruning and supernet training.
We also propose two novel methods, {\simpleTopK} and {\iterativeZeroout}, to automatically learn the optimal
layer combinations for the subnets through the training process.
Furthermore, we combine different training methods including layer dropout and the
sandwich rule to achieve better overall performance.
The experimental results on \textit{Librispeech} and \textit{TED-LIUM-v2} show that for each size, models trained using our approach can match
or slightly outperform models trained individually, and largely
outperform the models trained with auxiliary loss.
This shows that our training scheme can significantly reduce training redundancy
while preserving model performance.
For future work, the proposed approaches can be extended by using finer pruning granularity
for subnets such as the attention head in MHSA, the dimension of feed-forward layer, etc.
as in \cite{jiang2023microsoftpruning}.
Applying in-place knowledge distillation to transfer the knowledge from
supernet to subnets can also give potential improvements as reported in \cite{wu2021sparsity,aagaraja2021collaborative}.

\section{Acknowledgements}
This work was partially supported by the project RESCALE within the
program \textit{AI Lighthouse Projects for the Environment, Climate,
Nature and Resources} funded by the Federal Ministry for the
Environment, Nature Conservation, Nuclear Safety and Consumer
Protection (BMUV), funding ID: 67KI32006A.
\let\normalsize\small\normalsize
\let\OLDthebibliography\thebibliography
\renewcommand\thebibliography[1]{
        \OLDthebibliography{#1}
        \setlength{\parskip}{-0.31pt}
        \setlength{\itemsep}{1pt plus 0.07ex}
}
\bibliographystyle{IEEEtran}
\bibliography{mybib}

\end{document}